\documentclass[conference]{IEEEtran}

\IEEEoverridecommandlockouts
\usepackage{algorithm}
\usepackage{algorithmic}
\usepackage{comment}
\usepackage{color}
\usepackage{colortbl}
\usepackage{tabu}
\usepackage{soul}
\usepackage{courier}
\usepackage{multirow}
\usepackage{array}
\usepackage{stfloats}
\usepackage{hhline}
\usepackage{float}
\usepackage{subfig}
\usepackage{physics}
\usepackage{nccmath}
\usepackage{setspace}
\usepackage{booktabs}
\usepackage{pifont}
\usepackage{mwe}
\usepackage{cite}
\usepackage{amsmath,amssymb,amsfonts}
\usepackage{graphicx}
\usepackage{textcomp}
\usepackage{xcolor}
\usepackage[utf8]{inputenc} 
\usepackage{kotex} 

\usepackage{xspace}
\def\BibTeX{{\rm B\kern-.05em{\sc i\kern-.025em b}\kern-.08em
    T\kern-.1667em\lower.7ex\hbox{E}\kern-.125emX}}
\begin{document}
\newcommand{\BfPara}[1]{{\noindent\bf#1.}\xspace}
\newcommand\mycaption[2]{\caption{#1\newline\small#2}}
\newcommand\mycap[3]{\caption{#1\newline\small#2\newline\small#3}}

\title{Stable Marriage Matching for Traffic-Aware Space-Air-Ground Integrated Networks: \\ A Gale-Shapley Algorithmic Approach}

\author{
\IEEEauthorblockN{$^{\dag}$Hyunsoo Lee, $^{\dag}$Haemin Lee, $^{\ddag}$Soyi Jung, and $^{\dag}$Joongheon Kim}
\IEEEauthorblockA{$^{\dag}$Department of Electrical and Computer Engineering, Korea University, Seoul, Republic of Korea}
\IEEEauthorblockA{$^{\ddag}$School of Software, Hallym University, Chuncheon, Republic of Korea \\
E-mails: \texttt{hyunsoo@korea.ac.kr}, \texttt{haemin2@korea.ac.kr}, \texttt{sjung@hallym.ac.kr}, \texttt{joongheon@korea.ac.kr}}
}

\maketitle

\begin{abstract}
In keeping with the rapid development of communication technology, a new communication structure is required in a next-generation communication system. In particular, research using High Altitude Platform (HAP) or Unmanned Aerial Vehicle (UAV) in existing terrestrial networks is active. In this paper, we propose matching HAP and UAV using the Gale-Shapley algorithm in a relay communication situation. The numerical simulation results demonstrate that applying the Gale-Shapley algorithm shows superior performance compared to random matching.
\end{abstract}


\section{Introduction}

With the rapid development of 5G and Beyond 5G communication technology, sophisticated communication frameworks such as satellite-air-ground integrated network (SAGIN) have emerged as well as radio propagation methods. In \cite{iturf2438}, it analyzed the broadband high altitude platform (HAP) specific applications capacity demand and spectrum needs operating in the fixed services. In addition, the 3rd Generation Partnership Project (3GPP) published a technical report about New Radio (NR) to support Non-Terrestrial Networks (NTN) in \cite{3gpp38811}. Next-generation networks including HAP or unmanned aerial vehicles (UAVs) not only spend much less cost than building new terrestrial infrastructure but also have the potential to support higher data rates by using millimeter-wave wireless communications~\cite{tvt202108jung,6955961}. For instance, \cite{iotj17kim} proved that millimeter-wave technology could provide high rates for visual data delivery and high directionality for spatial reuse in edge computing networks. 
Meanwhile, relay communication in SAGIN can have trouble on account of the fast mobility of UAVs and the long distance between UAVs and HAPs~\cite{tvt202106jung,tvt201905shin,jsac201811dao,9467353}. Therefore, it becomes an essential issue for HAP to match which UAV to select. In addition to considering the distance, HAP needs to provide services to as many users as possible on the ground to satisfy ground users, so a method for finding an optimal matching in a given environment is needed. 

Motivated by the above considerations, in this paper we conduct the simulation to evaluate the feasibility of using the Gale-Shapley algorithm to match HAPs and UAVs to support services for on-the-ground user equipment (UE). In order to find the optimal matching, Gale-Shapley's algorithm is applied, which can always guarantee stable matching. As far as we know, there have been no studies that apply the Gale-Shapley algorithm between HAPs and UAVs. To put it briefly, the main contributions of the paper can be summarized as follows. 
\begin{itemize}
    \item This paper applied Gale-Shapley's stable marriage matching algorithm pairing between HAPs and UAVs.\
    \item We consider both the path loss calculated from distance and the number of ground users served by UAVs. \
    \item The performance of the proposed scheme is evaluated through simulations. We prove that our proposed algorithm outperforms that without matching.
\end{itemize}
The rest of this paper is organized as follows.
Sec.~\ref{sec:prelim} introduces the backgrounds of our proposed algorithm.
Sec.~\ref{sec:algo} presents the details of the proposed matching algorithm for HAPs and UAVs networks.
Sec.~\ref{sec:perf} evaluates the performance, and then, Sec.~\ref{sec:conclusion} concludes this paper.

\section{Preliminaries}\label{sec:prelim}
\subsection{Related Work}
Satellite network has been actively studied in recent years, due to the advent of next generation communication system.
In \cite{tb20hu}, it conducted a study to optimize the service of each beam, using a multi-objective reinforcement learning method to solve the beam hopping problem in a multi-beam satellite system. In \cite{jsac18du}, it carried out a study of secure communication in the situation where the satellite-terrestrial network was combined with the existing legacy network by mounting antennas in the satellite and terrestrial base stations in the mmWave frequency band.
HAPs are also often integrated with other networks. In \cite{access19zakaria}, a study was conducted to extend the joint transmission coordinated Multipoint (JT-CoMP)  method not only in the terrestrial cellular system but also to the HAP system using a phased array antenna. In \cite{vtc04pasquale}, it dealt with the network in which HAP supports dual-band connectivity in the Satellite-HAP-Terrestrial Network. In this scenario, HAP not only communicates directly with terrestrial users in C-band but also provides backhaul service through Ka-band.
 
Gale-Shapely Algorithm is the method that always guarantees stable matching. This Nobel Prize-winning method is a way to find a solution for the stable matching problem in economics and mathematics. However, it is also widely used in computer science and the matching of communication problems. In \cite{wcnc16chang}, it applied the Gale-Shapley algorithm to connect the device-to-device (D2D) pairs and cellular users to deal with the stringent interference~\cite{ton201608kim,jsac201806choi,mobicom13demo}. In \cite{wc16xiao}, it investigated a belief-based stable marriage game to analyze social-aware D2D communication. In this method, each user equipment can establish a function about belief to build a reliable social relationship in D2D communication in the social network. The Gale-Shapley algorithm is also applied in \cite{icccn17chowdhury} to solve channel assignment problems in cognitive radio networks.

\subsection{Network Model -- Satellite-HAPs-UAV Integrated Network}
\begin{figure} 
    \centering
    \includegraphics[scale=0.65]{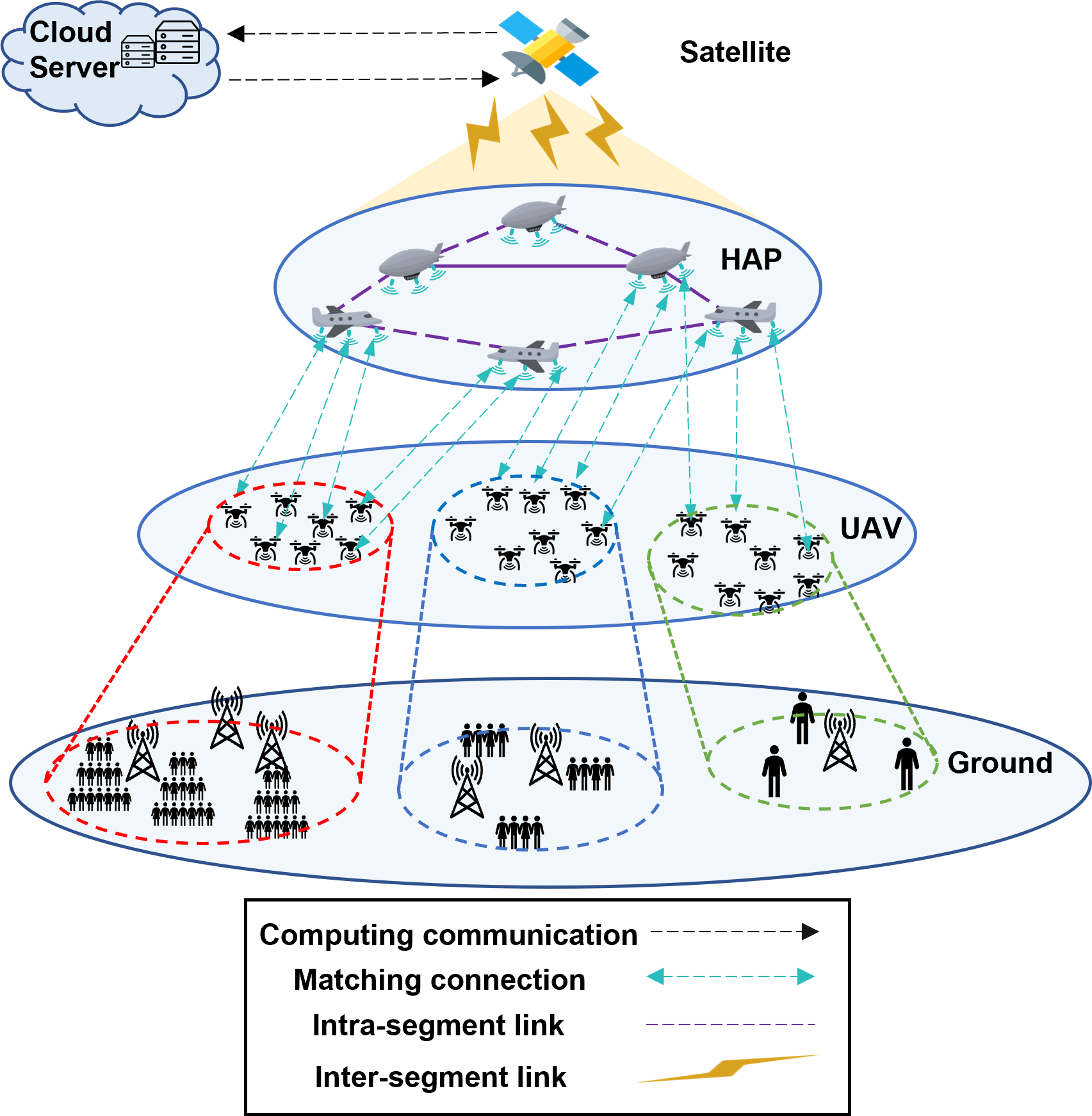} 
    \caption{HAPs-UAVs network scenario}
    \label{fig: Network model}
\end{figure}

We consider space-air integrated network consists of low earth orbit (LEO) satellite, HAPs, and UAVs described in Fig.~\ref{fig: Network model}. Each UAV is connected to HAP, and HAP also connects with the LEO satellite.
  Denote $U =\left\{ u_1, u_2, \cdots, u_m \right\} $ the set of UAVs. The set of the HAPs are given by $H=\left\{ h_1, h_2, \cdots, h_n \right\}$. Since HAP's antenna and UAV need to be matched, HAP's antenna is denoted by $H^i=\left\{ h_1^1, h_1^2, \cdots, h_n^i\right\}$ Matching $M(h_n^i,u_m)$ indicates the match between the antenna of HAP $h_n^i$ and the UAV $u_m$. 

\subsection{Communication Model}
\begin{table}[ht]
\begin{tabular}{|
>{\columncolor[HTML]{FFFFFF}}l |
>{\columncolor[HTML]{FFFFFF}}l |
>{\columncolor[HTML]{FFFFFF}}l |
>{\columncolor[HTML]{FFFFFF}}l |
>{\columncolor[HTML]{FFFFFF}}l |}
\hline
\textbf{Symbol}            & \multicolumn{4}{l|}{\cellcolor[HTML]{FFFFFF}\textbf{Description}}                                   \\ \hline
$U, H, M$         & \multicolumn{4}{l|}{\cellcolor[HTML]{FFFFFF}The set of UAVs, HAPs and matchings}       \\ \hline
$h_n^i$             & \multicolumn{4}{l|}{\cellcolor[HTML]{FFFFFF}The antenna of each HAP $h_n$}                      \\ \hline
$u_m, h_n$        & \multicolumn{4}{l|}{\cellcolor[HTML]{FFFFFF}The specific UAV $u_m \in U$, HAP $h_n \in H$} \\ \hline
$M(h^i_n$, $u_m$) & \multicolumn{4}{l|}{\cellcolor[HTML]{FFFFFF}The match between $h_n^i$ and $u_m$}           \\ \hline
$f_c$             & \multicolumn{4}{l|}{\cellcolor[HTML]{FFFFFF}Radio frequency between HAP and UAV}           \\ \hline
$d$               & \multicolumn{4}{l|}{\cellcolor[HTML]{FFFFFF}Distance between each HAP and UAV}             \\ \hline
\end{tabular}
\caption{Key notations}
\end{table}

\begin{figure*}[t!] 
    \centering
    \includegraphics[scale=0.7]{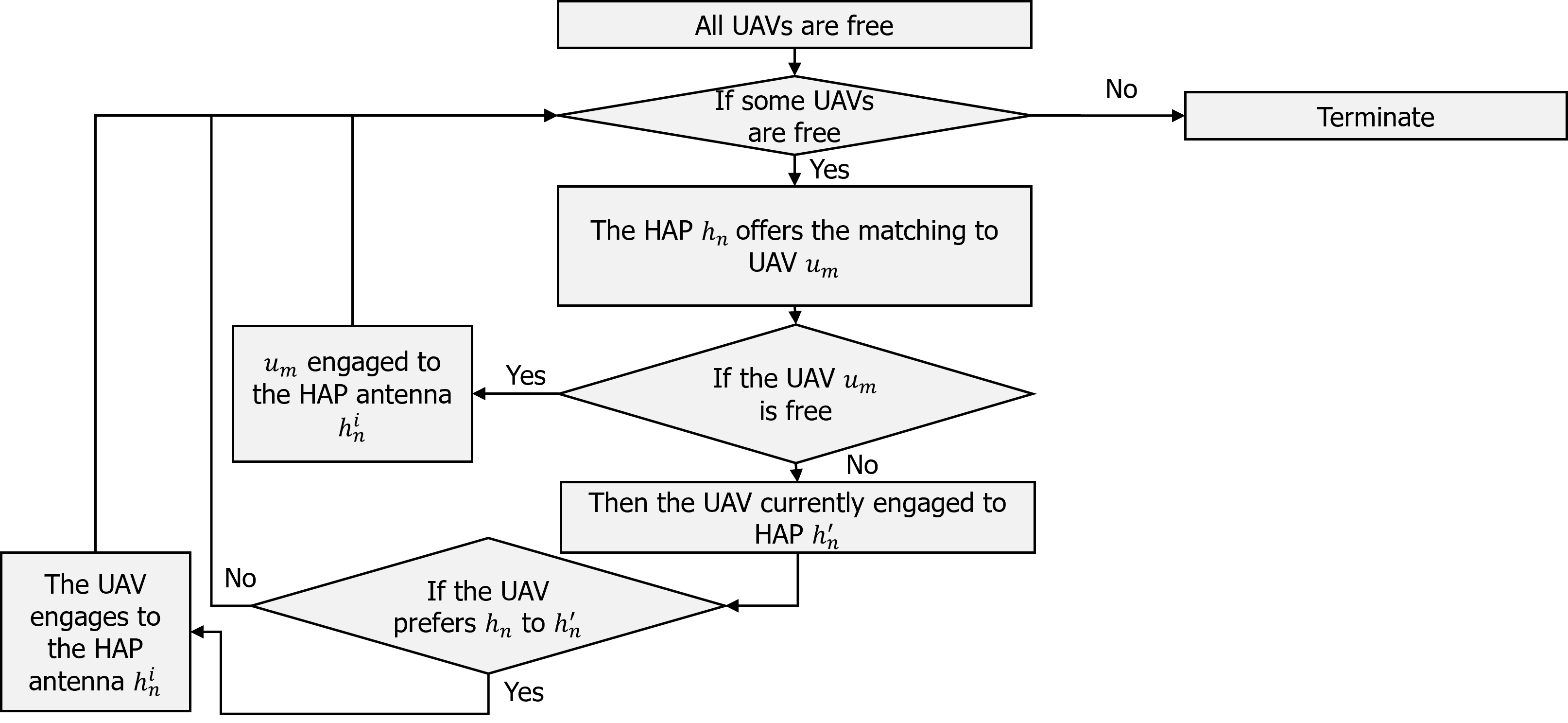}
    \caption{Gale-Shapley algorithm}
    \label{fig:GSA}
\end{figure*}
The key notations used in this paper are listed in Table I. Basically, the path-loss between UAV and HAP is measured and formulated in \cite{3gpp38811}, as follows. 

\begin{equation}
PL = PL_b + PL_g + PL_s 
\end{equation}
where $PL$ is the total path loss in dB, $PL_b$ is the basic path loss in dB, $PL_g$ is the attenuation due to atmospheric gasses in dB. According to \cite{jsac21jia}, we set $PL_g$ as 23 dB. $PL_s$ is the attenuation due to either ionospheric or tropospheric scintillation in dB. For latitudes between $\pm20^\circ$ and $\pm60^\circ$ of latitude, $PL_s$ = 0. 

The free space path loss (FSPL) over a distance $d$ in km and frequency $f_c$ in GHz is given by 
\begin{equation}
FSPL(d, f_c) = 92.45+20\log_{10}(f_c)+20\log_{10}(d)\mathrm. 
\end{equation}

The basic path loss in dB unit is modeled as  
\begin{equation}
PL_b = FSPL(d, f_c) + SF + CL(\alpha, f_c)\mathrm. 
\end{equation}
As stated in \cite{3gpp38811}, for the urban scenario, non-line-of-sight (NLOS) shadow fading follows a Gaussian distribution with zero mean and variance 6, and clutter loss is about 25.5 dB, respectively. We assume that the channel between HAPs and satellites is quasi-vacuum, and it is almost an ideal channel with additive white Gaussian Noise (AWGN). 

\section{Stable Matching for HAPs and UAVs}\label{sec:algo}

HAPs and UAVs should maintain stable matching in order to ensure smooth communication status. The matching between a HAP and a UAV follows the preference list based on two criteria.
First, HAP prefers the UAV with relatively low path loss. Since a UAV far away from the HAP consumes much power when matched, the HAP wants to select a UAV close to it.
HAP also prefers to match with a UAV which can provide services to more users. In other words, it is more advantageous for HAP to connect with a UAV that can service ten devices than a UAV that can provide a service to one device.

HAP $h_n$ with free antenna $h_n^i$ offers the matching to the UAV $u_m$. If the UAV $u_m$ is not matched with another HAP, a matching ($h_n^i, u_m$) will be created between the HAP and the UAV. 
On the other hand, if the UAV is matched with other HAP, then UAV $u_m$ compares the priority of the HAP $h_n$ and currently connected HAP $h'_n$. If the priority of $h'_n$ is higher than that of $h_n$, then UAV rejects the offer of new HAP $h_n$.

In contrast, if the priority of newly offered HAP $h_n$ is higher, UAV $u_m$ disconnects a current matching $({h'}_n^i, u_m)$, and matches with newly offered HAP $h_n$, building new match $(h_n^i, u_m)$. 
This process is repeated as long as there are unmatched UAVs. When all UAVs are matched, the process ends. The matching algorithm is described in Fig.~\ref{fig:GSA}.

\section{Performance Evaluation}\label{sec:perf}
\begin{figure}[t!] 
    \centering
    \includegraphics[scale=0.59]{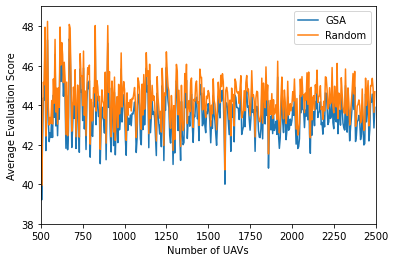} 
    \caption{Average score of random-based matching and Gale-Shapley algorithm-based matching}
    \label{fig: eval1}
\end{figure}
\subsection{Evaluation Setup}

The experiment assumes 100 to 500 HAPs operating at an altitude of 18 to 22 km in an urban area and 500 to 2500 UAVs operating at the height of 50 to 350 m above the ground. Each HAP has five antennas to connect with UAVs; therefore, maximum of 2500 matching is created when matching is over. UAVs prefer HAPs that can connect to themselves at lower path losses, and HAPs prefer UAVs that have lower path losses to themselves, as well as select UAVs that cover more users. Those preferences are reflected in the preference list of the Gale-Shapley algorithm. 

\subsection{Evaluation Results}

In this simulation, we devise a new score indicator to evaluate the algorithm, taking into account the path loss between UAV and HAP and the number of users served by the UAV. A lower score means that it is matched with a UAV with a low path loss while serving more users. Fig.~\ref{fig: eval1} indicates the average score of random-based matching and the Gale-Shapley algorithm-based matching. It is proved that the algorithm we applied shows better performance in all cases where matching is performed.

\begin{figure}[t!] 
    \centering
    \includegraphics[scale=0.59]{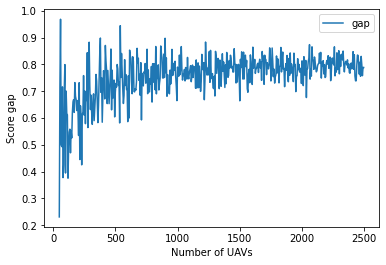} 
    \caption{Score gap between random-based matching and Gale-Shapley algorithm-based matching}
    \label{fig: eval2}
\end{figure}
Fig.~\ref{fig: eval2} shows the difference in scores between the two algorithms. As matching progresses, the algorithm we applied gradually shows stable and high performance compared to random matching.

\section{Concluding Remarks}\label{sec:conclusion}
In this paper, we conducted a study using the Gale-Shapley matching algorithm when connecting HAPs and UAVs. As mentioned above, there is a novelty in the idea of linking HAP and UAV with the Gale-Shapley algorithm, and the experimental results have proven that it performs better than random matching and an algorithm that only considers distance. We are going to consider various communication situations, such as interference between antennas in HAP or bandwidth allocation when connecting wireless communication networks between ground users and UAV in future work.

\section*{Acknowledgment}
This work was supported by Institute of Information \& communications Technology Planning \& Evaluation
(IITP) grant funded by the Korea government (MSIT) (No. 2021-0-00467,Intelligent 6G Wireless Access System). J. Kim and S. Jung are the corresponding authors of this paper (e-mails: joongheon@korea.ac.kr, sjung@hallym.ac.kr).

\bibliographystyle{IEEEtran}
\bibliography{ref_sagin,ref_opt,ref_quantum, ref_GSA,ref_aimlab}

\end{document}